\documentclass[fleqn,12pt,twoside]{article}
\usepackage[headings]{espcrc1}

\readRCS
$Id: espcrc1.tex,v 1.2 2004/02/24 11:22:11 spepping Exp $
\usepackage{graphicx}
\usepackage[figuresright]{rotating}

\newcommand {\bea}{\begin{eqnarray}}
\newcommand {\eea}{\end{eqnarray}}
\newcommand {\be}{\begin{equation}}
\newcommand {\ee}{\end{equation}}

\newcommand{\AmS}{{\protect\the\textfont2
  A\kern-.1667em\lower.5ex\hbox{M}\kern-.125emS}}

\title{Fermionic Quasiparticles in QCD at High Baryon Density}

\author{Thomas Sch{\"a}fer\address{Department of Physics, 
        North Carolina State University, Raleigh, NC 27695}}
       
\runtitle{Fermionic Quasiparticles in QCD}
\runauthor{T. Sch\"afer}

\begin{document}

% typeset front matter
\maketitle

\begin{abstract}
We study fermionic quasi-particles in QCD at very high baryon density.
In the normal quark matter phase unscreened magnetic gluon exchanges
lead to non-Fermi liquid behavior. Non-Fermi liquid effects manifest 
themselves in low energy Green functions that depend on logarithms
and fractional powers of energy. In the superfluid phase there is an 
energy gap for fermionic excitations. Quark mass effects can cause 
the energy gap to vanish. Gapless fermions in the color flavor 
locked phase cause an instability towards a state with a non-zero
supercurrent.
\end{abstract}

%%%%%%%%%%%%%%%%%%%%%%%%%%%%%%%%%%%%%%%%%%%%%%%%%%%%%%%%%%%%%%%%%%%%%%%
\section{Non-Fermi liquid behavior}
\label{sec_nfl}
%%%%%%%%%%%%%%%%%%%%%%%%%%%%%%%%%%%%%%%%%%%%%%%%%%%%%%%%%%%%%%%%%%%%%%%

%%%%%%%%%%%%%%%%%%%%%%%%%%%%%%%%%%%%%%%%%%%%%%%%%%%%%%%%%%%%%%%%%%%%%%%%%
\begin{figure}[t]
\begin{minipage}{6cm}
\vspace*{1cm}
\includegraphics[width=5cm]{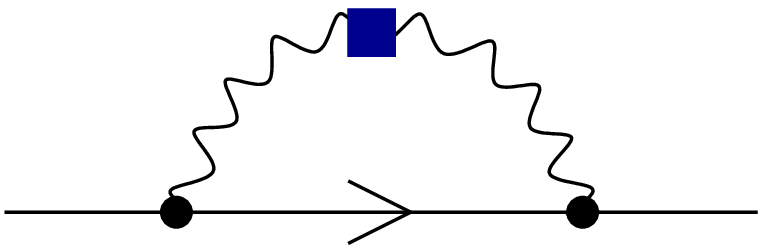}
\end{minipage}\begin{minipage}{9cm}
\includegraphics[width=9cm]{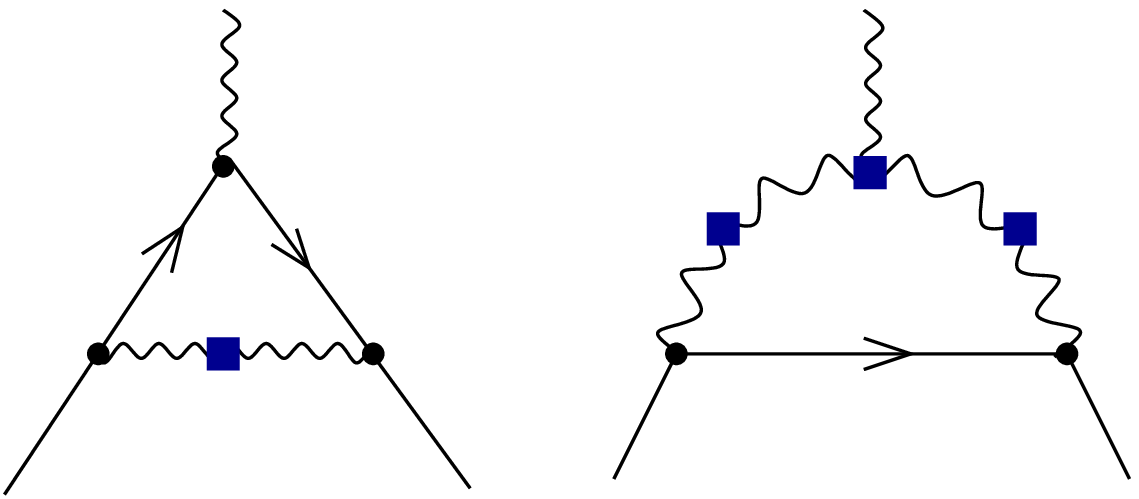}
\end{minipage}
\caption{One loop corrections to the fermion propagator and the 
quark gluon vertex in dense QCD. The solid squares denote hard 
dense loop insertions.}
\label{fig_sdl}
\end{figure}
%%%%%%%%%%%%%%%%%%%%%%%%%%%%%%%%%%%%%%%%%%%%%%%%%%%%%%%%%%%%%%%%%%%%%%%%%

 At high baryon density the relevant degrees of freedom are 
particle and hole excitations which move with the Fermi 
velocity $v$. Since the momentum $p\sim v\mu$ is large, 
typical soft scatterings cannot change the momentum by very 
much and the velocity is approximately conserved. An effective 
field theory of particles and holes in QCD is given by 
\cite{Hong:2000tn}
\be
\label{l_hdet}
{\cal L} =\psi_{v}^\dagger \left(iv\cdot D
   - \frac{1}{2p_F}D_\perp^2 \right) \psi_{v}
   + {\cal L}_{HDL} 
   -\frac{1}{4}G^a_{\mu\nu} G^a_{\mu\nu}+ \ldots ,
\ee
where $v_\mu=(1,\vec{v})$. The field $\psi_v$ describes particles 
and holes with momenta $p=\mu(0,\vec{v})+l$, where $l\ll\mu$. We will 
write $l=l_0+l_{\|}+l_\perp$ with $\vec{l}_{\|}=\vec{v}(\vec{l}
\cdot \vec{v})$ and $\vec{l}_\perp = \vec{l}-\vec{l}_{\|}$. 
At energies below the screening scale $g\mu$ hard dense loops
have to be resummed. The generating functional for hard dense loops 
in gluon $n$-point functions is given by \cite{Braaten:1991gm}
\be 
\label{S_hdl}
{\cal L}_{HDL} = -\frac{m^2}{2}\sum_v \,G^a_{\mu \alpha} 
  \frac{v^\alpha v^\beta}{(v\cdot D)^2} 
G^b_{\mu\beta},
\ee
where $m^2=N_f g^2\mu^2/(4\pi^2)$ is the dynamical gluon mass and 
the sum over patches corresponds to an average over the direction 
of $\vec{v}$. 

 The hard dense loop action describes static screening of electric 
fields and dynamic screening of magnetic modes. Since there is no 
screening of static magnetic fields low energy gluon exchanges are 
dominated by magnetic modes. The resummed transverse gauge boson 
propagator is given by
\be
\label{d_trans}
D_{ij}(k) = \frac{\delta_{ij}-\hat{k}_i\hat{k}_j}{k_0^2-\vec{k}^2+
i\eta |k_0|/|\vec{k}|} ,
\ee
where $\eta=\frac{\pi}{2}m^2$ and we have assumed that $|k_0|<|\vec{k}|$. 
We observe that the gluon propagator becomes large in the regime 
$|\vec{k}|\sim (\eta k_0)^{1/3}\gg k_0$. This leads to an unusual scaling 
behavior of Green functions in the low energy limit. Consider a 
generic Feynman diagram and scale all energies by a factor $s$.
Because of the functional form of the gluon propagator in the 
Landau damped regime gluon momenta scale as $|\vec{k}|\sim s^{1/3}$. 
This implies that the gluon momenta are much larger than the 
gluon energies. The quark dispersion relation is $k_0\simeq 
k_{||}+k_\perp^2/(2p_F)$. The only way a quark can emit a very 
spacelike gluon and remain close to the Fermi surface is if the 
gluon momentum is transverse to the Fermi velocity. We find
\be 
k_0 \sim s, \hspace{0.5cm}
k_{||}\sim s^{2/3},\hspace{0.5cm}
k_\perp \sim s^{1/3},
\ee
and $k_0\ll k_{||}\ll k_\perp$. These scaling relations have many 
interesting consequences. As an example we consider the fermion self 
energy, see Fig.~\ref{fig_sdl}. The one-loop self energy is  
\be
\Sigma(p) \sim \int dk_0
\int dk_\perp^2 \ \frac{k_\perp}{k_\perp^2+i\eta k_0}
 \int dk_{||} \  \frac{\Theta(p_0+k_0)}{k_{||}+p_{||}
          -(k_\perp^2+p_\perp^2)/(2p_F)+i\epsilon}
 \sim p_0\log(p_0).
\ee
A more careful calculation gives \cite{Ipp:2003cj}
\be
\label{sig}
\Sigma(\omega) = \frac{g^2}{9\pi^2} \left[ 
  \omega \log\left( \frac{4\sqrt{2}m}{\pi|\omega|} \right)
  +\omega + i\frac{\pi}{2}|\omega| \right].
\ee
There are no corrections of the form $g^{2n}\log^n(\omega)$
\cite{Schafer:2004zf}. Higher order corrections involve fractional 
powers $(\omega/m)^{1/3}$. Eq.~(\ref{sig}) shows that cold quark matter 
is not a Fermi liquid. The Fermi velocity vanishes on the Fermi surface 
and the specific heat scales as $T\log(T)$. 

%%%%%%%%%%%%%%%%%%%%%%%%%%%%%%%%%%%%%%%%%%%%%%%%%%%%%%%%%%%%%%%%%%%%%%%
\section{CFL Phase}
\label{sec_cfl}
%%%%%%%%%%%%%%%%%%%%%%%%%%%%%%%%%%%%%%%%%%%%%%%%%%%%%%%%%%%%%%%%%%%%%%%

 If the temperature is sufficiently low quark matter is 
expected to become a color superconductor and quarks acquire 
a gap due to pairing near the Fermi surface. In the following 
we will concentrate on the color flavor locked (CFL) phase 
which is the ground state of three flavor quark matter at very 
high baryon density. Our starting point is the effective theory 
of the CFL phase derived in \cite{Casalbuoni:1999wu,Kryjevski:2004jw}.
The effective lagrangian contains Goldstone boson fields $\Sigma$ 
and baryon fields $N$. The meson fields arise from chiral symmetry 
breaking in the CFL phase. The leading terms in the effective theory 
is
\be
\label{l_mes}
{\cal L} =  \frac{f_\pi^2}{4} {\rm Tr}\left(
  \nabla_0 \Sigma \nabla_0 \Sigma^{\dagger} - v_\pi^2
  \vec{\nabla} \Sigma \vec{\nabla}\Sigma^\dagger \right).
\ee
Baryon fields originate from quark-hadron complementarity. The
effective lagrangian is 
\bea
\label{l_bar}
 {\cal L} &=& 
 {\rm Tr}\left(N^\dagger iv^\mu D_\mu N\right)
 - D{\rm Tr} \left(N^\dagger v^\mu\gamma_5
               \left\{ {\cal A}_\mu,N\right\}\right)
 - F{\rm Tr} \left(N^\dagger v^\mu\gamma_5
               \left[ {\cal A}_\mu,N\right]\right)
  \nonumber \\
 & &  \mbox{} + \frac{\Delta}{2} \left\{
     \left( {\rm Tr}\left(N_LN_L \right)
   - \left[ {\rm Tr}\left(N_L\right)\right]^2 \right)
   - \left( {\rm Tr} \left(N_RN_R \right)
   - \left[ {\rm Tr}\left(N_R\right)\right]^2 \right)
     + h. c.  \right\},
\eea
where $N_{L,R}$ are left and right handed baryon fields in the adjoint 
representation of flavor $SU(3)$, $v^\mu=(1,\vec{v})$ is the Fermi 
velocity, and $\Delta$ is the superfluid gap. We can think of $N$
as being composed of a quark and a diquark field, $N_L\sim q_L
\langle q_Lq_L\rangle$. The interaction of the baryon field with 
the Goldstone bosons is dictated by chiral symmetry. The covariant 
derivative is given by $D_\mu N=\partial_\mu N +i[{\cal V}_\mu,N]$. 
The vector and axial-vector currents are
\be
 {\cal V}_\mu = -\frac{i}{2}\left\{
  \xi \partial_\mu\xi^\dagger +  \xi^\dagger \partial_\mu \xi
  \right\}, \hspace{1cm}
{\cal A}_\mu = -\frac{i}{2} \xi\left(\partial_\mu
    \Sigma^\dagger\right) \xi ,
\ee
where $\xi$ is defined by $\xi^2=\Sigma$. The low energy constants 
$f_\pi,v_\pi,D,F$ can be calculated in perturbative QCD. Symmetry 
arguments can be used to determine the leading mass terms in the 
effective lagrangian. Bedaque and Sch\"afer observed that $X_L=
MM^\dagger/(2p_F)$ and $X_R=M^\dagger M/(2p_F)$ act as effective 
chemical potentials and enter the theory like the temporal components 
of left and right handed flavor gauge fields \cite{Bedaque:2001je}. 
We can make the effective lagrangian invariant under this symmetry 
by introducing the covariant derivatives
\bea
\label{V_X}
 D_0N &=& \partial_0 N+i[\Gamma_0,N], \hspace{0.5cm}
 \Gamma_0 = -\frac{i}{2}\left\{
  \xi \left(\partial_0+ iX_R\right)\xi^\dagger +
  \xi^\dagger \left(\partial_0+iX_L\right) \xi
  \right\}, \\
\nabla_0\Sigma &=& \partial_0\Sigma+iX_L\Sigma-i\Sigma X_R.
\eea
Using eqs.~(\ref{l_bar}-\ref{V_X}) we can calculate the dependence
of the gap in the fermion spectrum on the strange quark mass. For
$m_s=0$ there are 8 quasi-particles with gap $\Delta$ and one
quasi-particle with gap $2\Delta$. As $m_s$ increases some of 
the gaps decrease. The gap of the lowest mode is approximately
given by $\Delta=\Delta_0-3\mu_s/4$ where $\mu_s=m_s^2/(2p_F)$
and $\Delta_0$ is the gap in the chiral limit. 

%%%%%%%%%%%%%%%%%%%%%%%%%%%%%%%%%%%%%%%%%%%%%%%%%%%%%%%%%%%%%%%%%%%%%%%%%
\begin{figure}
\begin{center}
\includegraphics[width=10cm]{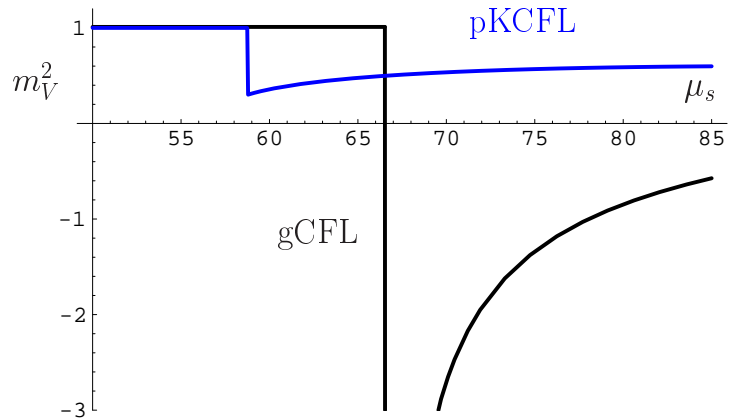}
\vspace*{-1cm}
\end{center}
\caption{Screening mass in the CFL phase as a function of the 
effective chemical potential $\mu_s=m_s^2/(2p_F)$. The screening 
mass is defined as the second derivative of the energy density 
with respect to an isospin or hypercharge current. The blue 
and black curves show the result in the gapless CFL phase with 
(pKCFL) and without (gCFL) a supercurrent.  }
\label{fig_kst}
\end{figure}
%%%%%%%%%%%%%%%%%%%%%%%%%%%%%%%%%%%%%%%%%%%%%%%%%%%%%%%%%%%%%%%%%%%%%%%%%

 For $\mu_s>4\Delta_0/3$ the system contains gapless fermions
interacting with light or even massless Goldstone bosons. This
situation is superficially similar to the normal phase discussed
in Sect.~\ref{sec_nfl}, but this is not the case. The gapless
CFL phase is unstable with respect to the formation of non-zero
currents \cite{Huang:2004bg}. This can be seen by considering 
the dispersion relation in the presence of a hypercharge or 
isospin current. The dispersion relation of the lowest mode
is \cite{Kryjevski:2005qq,Schafer:2005ym}
\be
\label{disp_ax}
\omega_l = \Delta_0 +\frac{l^2}{2\Delta_0}-\frac{3}{4}
  \mu_s -\frac{1}{4}\vec{v}\cdot\vec{\jmath}_K,
\ee
where $l$ is the momentum relative to the Fermi surface and 
$\jmath_K$ is the current. The energy relative to the CFL phase 
is the kinetic energy of the current plus the energy of occupied 
gapless modes
\be
\label{efct}
{\cal E} = \frac{1}{2}v_\pi^2f_\pi^2\vec{\jmath}_K^2 
 + \frac{\mu^2}{\pi^2}\int dl \int
 \frac{d\Omega}{4\pi} \;\omega_l \theta(-\omega_l) .
\ee
The energy functional can develop a minimum at non-zero $\jmath_K$ 
because the current lowers the energy of the fermions near one of 
the pole caps on the Fermi surface. Introducing the dimensionless 
variables $x=\jmath_K/(a\Delta)$ and $h=(3\mu_s-4\Delta)/(a\Delta)$
we can write
\be
\label{efct_x}
{\cal E} = Cf_h(x), \hspace{0.5cm}
 f_h(x) = x^2-\frac{1}{x}\left[
   (h+x)^{5/2}\Theta(h+x) - (h-x)^{5/2}\Theta(h-x) \right] ,
\ee
where $C$ and $a$ are numerical constants. The functional eq.~(\ref{efct_x}) 
was analyzed in \cite{Son:2005qx,Kryjevski:2005qq,Schafer:2005ym}.
There is a critical chemical potential $\mu_s= (4/3+ah_{crit}/3)\Delta$
above which the groundstate contains a non-zero supercurrent $\jmath_K$.
This current is canceled by a backflow of gapless fermions. The 
screening mass $m_V^2=(\partial^2{\mathcal E})/(\partial\jmath_K^2)$
is shown in Fig.~\ref{fig_kst}. Without the supercurent an instability 
occurs for $\mu_s= 4\Delta/3$, but the instability is resolved by 
a non-zero current. The new phase is analogous to $p$-wave pion
condensates at lower densities because the current is carried by 
Goldstone kaons in the CFL phase and the instability is caused 
by the $p$-wave interaction between kaons and fermions.

\end{document}